\documentclass{article}

\usepackage{ijcai26}

\usepackage{times}
\usepackage{soul}
\usepackage{url}
\usepackage[hidelinks]{hyperref}
\usepackage[utf8]{inputenc}
\usepackage[small]{caption}
\usepackage{graphicx}
\usepackage{amsmath}
\usepackage{amsthm}
\usepackage{booktabs}
\usepackage{fancyvrb}
\usepackage[switch]{lineno}
\usepackage{multirow}
\usepackage{microtype}
\usepackage{tikz}
\usepackage{macros}
\usepackage{listings}
\usepackage{inconsolata}
\usepackage[ruled,vlined,linesnumbered]{algorithm2e}
\usepackage[inline]{enumitem}
\usepackage{pgfplots}
\usepackage{comment}
\usepackage[table]{xcolor}

\newcommand\pr[1]{{\color{teal}[R: #1]}}

\newtheorem{theorem}{Theorem}

\title{Large Lemma Miners:\\ Can LLMs Do Induction
Proofs for Hardware?\\
(Accepted to IJCAI'26)}

\author{
Romy Peled$^{1,2}$
\and
Daniel Kroening$^{2,3}$\and
Michael Tautschnig$^{2,4}$\And
Yakir Vizel$^{1}$\\
\affiliations
$^1$Technion – Israel Institute of Technology \\
$^2$Amazon\\
$^3$University of Oxford \\
$^4$Queen Mary University of London\\
\emails
romypel@amazon.com,
dkr@amazon.com,
tautschn@amazon.at,
yvizel@cs.technion.ac.il
}

\begin{document}

\maketitle

\begin{abstract}

Large Language Models (LLMs) have shown potential for solving mathematical tasks.
%
We show that LLMs can be utilized to generate proofs by induction for hardware verification and thereby replace some of the manual work 
done by Formal Verification engineers and deliver value to industry.
We present a neurosymbolic approach that includes two prompting frameworks to generate candidate invariants, which are checked using a formal symbolic tool. 
Our results indicate that with sufficient reprompting, LLMs are able to generate inductive arguments for mid-size open-source RTL designs. For 
$90\%$ of our problem set, at least one of the prompt setups succeeded in producing a provably correct inductive argument. 
\end{abstract}

\section{Introduction}

Large Language Models (LLMs) have attracted enormous attention, research effort, and financial investment for their potential to perform mathematical reasoning. The International Math Olympiad (IMO), in particular, has served as a high-profile testbed for demonstrating advanced mathematical reasoning capabilities of state-of-the-art (SOTA) LLMs. Remarkably, SOTA models have matched the performance of leading contestants at the 2025 IMO, with Google DeepMind achieving an official gold-medal standard~\footnote{\url{https://tinyurl.com/5n7abr7h}}
.
It is therefore only natural to consider LLMs as a tool that can assist in complex verification tasks, which currently rely on humans doing mathematical reasoning.

Formal Verification (FV) is a crucial step in the design cycle of hardware, supported by decades of research and substantial industrial investment. Modern FV tools rely on well-established model checking algorithms~\cite{bitlevel}. Yet, the application of FV tools to large industrial designs remains a challenge. This is mainly due to the fact that most FV tools operate at the ``bit-level''.
This prevents these tools from constructing a high-level mathematical proof that is agnostic to the number of bits in the design. Constructing such proofs is how a human mathematician approaches such problems.
As a result, verifying complex properties that cover meaningful functionality of the design requires extensive manual effort, which is usually performed by experienced FV engineers.

One of the main FV methods is \emph{proof by induction}. 
This method requires the use of auxiliary lemmas that capture key intermediate facts. 
These lemmas must first be proven correct and can then be used to construct an inductive argument establishing the property of interest.

In this paper, we set out to answer the following question: \emph{can Large Language Models (LLMs) be used to mine such auxiliary lemmas and construct inductive arguments for proving complex properties?} Such a capability has far-reaching industrial implications, as it may greatly increase the applicability of formal verification for hardware designs.



Our initial experiments with LLMs have exhibited a well-known pattern: the state-of-the-art LLMs produce many answers, and many of the lemmas generated by the LLMs are indeed correct inductive arguments.  However, the LLMs also hallucinate: in addition to the correct lemmas, we obtain lemmas that do not even parse, are not inductive, or do not prove the property we are interested in.

We therefore propose a \emph{neurosymbolic} approach that combines the LLM with a formal checking framework that sifts through the LLM's output to find lemmas that form a valid formal proof.
Our lemma mining framework has two main components: \begin{enumerate*}[label=(\roman*)]
    \item A single Non-Agentic \emph{Few-Shot} prompt and an \emph{Agentic} LLM prompting setup that are responsible for suggesting candidate lemmas that can serve as inductive arguments; and 
    \item a formal reasoning algorithm that finds among the outputs of the LLMs a subset of useful lemmas that form an inductive argument for the given verification task.
\end{enumerate*}
Then, to evaluate our framework, we use a dataset consisting of 109 verification tasks. A verification task consists of a \systemverilog design and a property to be verified. We evaluate our LLM-based framework on this dataset and show 
that for $90\%$ of these verification tasks, at least one of our prompting setups produces a provably correct inductive argument. In addition, to demonstrate the effectiveness of our LLM-based framework and to probe the extent to which it can produce nontrivial inductive arguments, we identify verification tasks that are challenging for SOTA tools and evaluate our approach against them. Our results show that our LLM-based approach can solve instances that SOTA tools fail to solve.

    Our work makes the following contributions:
    \begin{enumerate}
        \item We present the use case of neurosymbolic lemma generation to construct formal proofs for hardware verification.
        \item We implement a lemma mining framework, which uses two LLM prompting setups, one of them Agentic, to generate candidate lemmas, and detail the methods used to evaluate their effectiveness.
        \item We provide an empirical study of our suggested framework that highlights both the potential and the current limitations of our approach.

    \end{enumerate}

\section{Related Work}








There is a large body of work on the application of Artificial Intelligence (AI) to mathematics~\cite{advancing_math_human_intuition,alpha_tensor,
transformers_lyapunov,ml_invariants_arithmetic_curves,predicting_root_numbers,euler_factors_transformers,mathematical_discoveries_program_search}. In particular, many LLM-based approaches are used in mathematics and the generation of formal proofs for interactive theorem provers~\cite{draft_sketch_prove,lean_star,study_in_sat_solving,proof_or_bluff,formal_mathematical_reasoning,llms_reasoning_survey,transformers_as_soft_reasoners,proof_writer,mixture_of_thought}. These works do not consider hardware.

The works in~\cite{can_chatgpt_support_sv,learning_invariants_using_llms,lemur,llms_proof_synth_rust} study the ability of LLMs to generate loop invariants or class invariants~\cite{DBLP:journals/corr/abs-2502-18917} for programming languages such as \textsc{C} and \textsc{Rust}. This can be viewed as somewhat similar to our goal of constructing inductive arguments for hardware proofs. However, these two tasks are conceptually different. Loop invariants are local to the loop and require the right state predicate that is preserved by the loop body. By contrast, inductive arguments for a hardware design are global, can include temporal modality/operators, and capture facts about all reachable states of a transition system. 
The use of Agentic setups 
as a mechanism to deal with incorrect LLM responses is commonplace for software engineering tasks~\cite{DBLP:journals/pacmpl/ZhangDWPK25,DBLP:journals/corr/abs-2404-18852}.

Machine learning has been applied for hardware proofs in~\cite{neural_model_checking}. However, this work 
trains a task-specific neural network rather than using an LLM, and produces formal guarantees in the form of ranking functions. 

LLMs have been applied to hardware designs in~\cite{llms_fv_rtl,assertllm,assertions_by_llms,verigen,SecV}. In~\cite{llms_fv_rtl}, an LLM is used to generate \sva (SVA) from RTL code; \cite{assertllm} propose a framework for capturing functional behavior in SVA from specification files containing waveform and natural language descriptions; \cite{DBLP:conf/date/KangLHSR25} evaluate LLMs on the generation of SVA assertions from natural language specifications or from RTL alone; \cite{assertions_by_llms} investigate the use of LLMs in generating hardware security assertions; \cite{verigen} fine-tune pretrained LLMs to produce syntactically correct \verilog code; and \cite{SecV} construct hardware common weakness enumeration (CWE) knowledge graphs to generate secure \verilog code. \cite{DBLP:conf/dac/Qayyum0A0D24} incrementally constructs invariants for design modules. 
Yet, none of these works use LLMs to produce proofs for hardware with respect to a given formal specification.

\section{Motivating Example}

The following is a short example in our dataset of a round-robin arbiter, taken from \texttt{v2c}~\cite{DBLP:conf/tacas/MukherjeeTK16}. Comments have been removed to save space.
\begin{lstlisting}[language=Verilog,
                   morekeywords={property,endproperty},
                   basicstyle=\scriptsize\ttfamily]
module main(clk,rst,ir0,ir1,ack0,ack1);
    input  clk, rst, ir0, ir1;
    output ack0, ack1;
    reg req0, req1, ack0, ack1, robin;

    task initialize; begin
	    ack0 = 0; ack1 = 0; robin = 0;
	    req0 = ir0; req1 = ir1;	
    end
    endtask

    always @ (posedge clk) begin
    if (rst) initialize;
    else begin
        if (~req0) ack0 <= 0;		
        else if (~req1) ack0 <= 1;		
        else if (~ack0 & ~ack1) ack0 <= ~robin;	
        else ack0 <= ~ack0;		
  
        if (~req1) ack1 <= 0;		
        else if (~req0) ack1 <= 1;		
        else if (~ack0 & ~ack1) ack1 <= robin;		
        else ack1 <= ~ack1;		
    
        if (req0 & req1 & ~ack0 & ~ack1)
          robin <= ~robin;	
          req0 <= ir0;
          req1 <= ir1;
      end     
  end         
 
  property prop;
    @(posedge clk) disable iff (rst) 
      (req1==1 && ack0==1 |-> ##1 ack1==1);
  endproperty
endmodule 
\end{lstlisting}
The specified property, while true, is not an inductive invariant. 

When executing the LLM-based framework presented in this paper on this example, all variants generate an inductive strengthening. Among others, the following lemma is generated:

\begin{lstlisting}[language=Verilog,
                   morekeywords={property,endproperty},
                   basicstyle=\scriptsize\ttfamily]
property lemma_1; 
@(posedge clk) disable iff (rst) 
  ~(ack0 && ack1); 
endproperty
\end{lstlisting}

The suggested lemma holds in the given design and forms an inductive strengthening for the specified property. That is, the following property is an inductive invariant:

\begin{lstlisting}[language=Verilog,
                   morekeywords={property,endproperty},
                   basicstyle=\scriptsize\ttfamily]
property lemma_prop; 
  lemma_1 and prop; 
endproperty
\end{lstlisting}

\section{Preliminaries}

\subsection{Safety Verification}

Without loss of generality we formalize the safety verification problem using reachability properties. Our approach is not limited to reachability and is applicable to the full safety fragment of LTL and SVA.

An RTL design $D$ and an LTL safety property $\varphi$ can be translated into a reachability problem
by compiling the RTL and the property into a transition system and a reachability property of the form $\mathbf{AG}\Prop$. 
A~transition system $T$ is a tuple~$(\Vars, \Init, \Tr)$, where $\Vars$ is a set of Boolean variables that defines the states of~$T$ (i.e., 
all valuations of $\Vars$), $\Init$~is a formula with variables in $\Vars$ defining the set
of initial states, and~$\Tr$ is a formula that has
free variables in $\Vars \cup \Vars'$, defining the transition
relation. A~state $s\in T$ is said to be reachable iff
there exists a state~$s_0\in\Init$ and $(s_i,s_{i+1})\in\Tr$ for $0\leq i < N$, such that $s = s_N$.
The property $\mathbf{AG}\Prop$ requires that $\Prop$, which is a formula over $\Vars$ representing the set of
``safe states'',
holds in all reachable states of $T$.

\begin{definition}[Inductive Invariant]\label{def:invar}
    Given a transition system~$T$, an \emph{inductive invariant} is a formula $\Inv$ satisfying: $\Init(\Vars) \limp Inv(\Vars)$ (initiation) and $\Inv(\Vars) \land \Tr(\Vars,\Vars') \limp \Inv(\Vars')$ (consecution).
\end{definition}
A transition system $T$ is \emph{safe} iff all reachable states in $T$ 
satisfy~$\Prop$. Equivalently, there exists a \emph{safe inductive invariant} $\Inv$ satisfying: $\Inv$ is an inductive invariant and $\Inv(\Vars)
  \limp \Prop(\Vars)$. 
A \emph{safety} verification problem is to decide whether a transition
system $T$ is safe or not, i.e., whether there exists a safe inductive invariant, or a counterexample that shows $\neg \Prop$ is reachable from an initial state.

In simple cases, $\Prop$ is already an inductive invariant.
Complex properties, however, are often not inductive and an inductive invariant $\Inv$ must be synthesized to prove them. However, model checking algorithms
~\cite{DBLP:books/daglib/0007403-2}, 
which are the backbone of most EDA tools, sometimes fail to synthesize $\Inv$ in a reasonable amount of time.

In such scenarios, manual intervention is required. An FV engineer can add auxiliary lemmas that form a safe inductive invariant, making the safety verification problem tractable for model checking algorithms.

\begin{definition}[Inductive Strengthening]
    Given a transition system $T = (\Vars, \Init, \Tr)$, let $L=\{\Lemma_1,\ldots,\Lemma_k\}$ be a set of properties. $L$ is an inductive strengthening for $\mathbf{AG}\Prop$ if $\Land\limits_{i=1}^k q_i\land\Prop$ is an inductive invariant.

 
\end{definition}

\subsection{Large Language Models}
We regard an LLM as a function that maps textual input to textual output. The input may take the form of a single prompt or a sequence of messages (i.e., a conversation context). While LLMs are fundamentally deterministic, they are typically used with deliberately randomized decoding strategies that allow the same input to produce different outputs across different queries. Hence, it is often highly effective to re-prompt the LLM with the same input in order to obtain varied responses and gather additional information, as demonstrated in numerous prior works~\cite{learning_invariants_using_llms,DBLP:journals/nature/FarquharKKG24,DBLP:journals/corr/abs-2406-15927}. Throughout this paper, when we query an LLM $k$ times with the same input---whether it is a single prompt in the Non-Agentic setup or a sequence of messages in the Agentic setup---we refer to this as obtaining $k$ sample responses for that input.



Two well-known prompting strategies for LLMs are Few-Shot and Chain-of-Thought (CoT) prompting. In Few-Shot, the input for the LLM contains several auxiliary examples in the form of $\langle question, answer\rangle$. These examples illustrate the expected output format and guide the LLM towards giving appropriate solutions.
Chain-of-Thought prompting extends this idea by additionally providing a sequence of reasoning steps for every $\langle question, answer\rangle$ example. This chain of reasoning  mimics the way a human would reason about the question and derive the answer.
Chain-of-Thought prompting has been shown to elicit 
reasoning traces from LLMs and to substantially improve their performance on reasoning tasks \cite{chain_of_thought_elicits_reasoning_neurips}. 

\section{Lemma-Mining Framework} 

In this section, we describe our lemma mining framework. We start by describing our prompting configurations that drive the LLM to generate candidate lemmas. 
We then briefly describe an algorithm that, given a design, property and set of candidate lemmas, finds an inductive strengthening for the property. 

\subsection{Prompting Setups}

To generate candidate lemmas with LLMs, we develop two configurations as described below. Throughout this section we consider a pair of an RTL design and a property, $(D,\varphi)$, for which we would like to construct an inductive strengthening.

\subsubsection{Single Prompt, Non-Agentic}

\begin{algorithm}[t]
\small
\caption{Non-Agentic}\label{alg:non_agentic}
\KwIn{A pair $(D,\varphi)$ of a \systemverilog design and a property, respectively; positive integers $k$ and $n$}
\KwOut{An inductive strengthening $L$}
    $S\gets\text{\texttt{GetFewShotPrompt}}(D,\varphi,k)$\;\label{alg:line:non_agentic_prompt}
    $C\gets\emptyset$\;
    \For {$i=1\ldots n$\label{alg:line:reprompt_begin}} {
        $C\gets C\cup \text{\texttt{LLM.Prompt}}(S)$\;\label{alg:line:reprompt_end}
    }
    $L\gets \lemmamine(D,\varphi,C)$\;\label{alg:line:non_agentic_ind}
    \Return {$L$}\;
\end{algorithm}


The prompt used in Algorithm~\ref{alg:non_agentic} (line~\ref{alg:line:non_agentic_prompt}) is few-shot. Note that it remains constant throughout the execution of the algorithm. To create the few-shot prompt, we construct a pool of Chain-of-Thought  responses (CoT pool). This pool consists of triples that include an RTL design, a property, and a set of lemmas that serve as an inductive strengthening w.r.t.~the property and design. Since this is a CoT, responses also include detailed reasoning as to why the given set of lemmas serves as an inductive strengthening. 

Given a pair $(D,\varphi)$, the constructed few-shot prompt for this pair has a fixed template followed by $k$ examples (i.e., triples) from the CoT pool described above, where $k$ is a parameter. The fixed template specifies the task, provides general guidelines, and defines the expected output format. 
Note that since only $k$ examples are chosen from the pool, it is desirable to choose those that can guide the LLM towards the best response. For this purpose, our framework embeds $(D,\varphi)$ and all entries in the pool using \textit{sentence transformers}~\cite{DBLP:conf/emnlp/ReimersG19}, and chooses the $k$ examples that are most similar to $(D,\varphi)$ based on the dot product of the computed embeddings.

Since the LLM may return different responses for the same prompt, we employ sampling (i.e., prompting the LLM multiple times with the same prompt, see Section~\ref{sec:experimental_setup}). The number of re-prompting iterations is determined by a parameter $n$ (lines~\ref{alg:line:reprompt_begin}--\ref{alg:line:reprompt_end}).
All LLM responses are collected in a set of candidate lemmas $C$ (line~\ref{alg:line:reprompt_end}). Then, the set $C$ is passed to \lemmamine (see Section~\ref{sec:find_ind}), which is an algorithm that attempts to return an inductive strengthening, if such 
exists in~$C$ (line~\ref{alg:line:non_agentic_ind}). 

\subsubsection{LLM Agent}

\begin{algorithm}[t]
\SetNlSkip{0.3em}
\small
\caption{Agentic}\label{alg:agentic}
\KwIn{A pair $(D,\varphi)$ of a \systemverilog design and a property, respectively; positive integers $k$ and $n$}
\KwOut{An inductive strengthening $L$}
    $S\gets\text{\texttt{GetFewShotPrompt}}(D,\varphi,k)$\;\label{line:alg:agentic_prompt}
    $C\gets\emptyset$, $All\gets\emptyset$\;
    \For {$i=1\ldots n$\label{alg:line:loop_start}} {
        $C\gets \text{\texttt{LLM}.Prompt}(S)$\;\label{alg:line:agentic_cand}
        $L\gets \lemmamine(D,\varphi,C)$\;\label{alg:line:agentic_ind}
        \If{$L\neq \emptyset$} {
            \Return {$L$}\;
        }
        $All\gets All\cup C$\;
        $L\gets \lemmamine(D,\varphi,All)$\;\label{alg:line:agentic_ind_all}
        \If{$L\neq \emptyset$} {
            \Return {$L$}\;
        }
        $S\gets \text{\texttt{Concat}}\;(S,\text{\texttt{GenerateRepairMsg}}(C))$\;\label{alg:line:repair}
    }
    \Return {$\emptyset$}\;
\end{algorithm}

In the Agentic setting of Algorithm~\ref{alg:agentic}, lemma generation is framed as an iterative interaction between an LLM and a verifier. Instead of generating one, static batch of candidate lemmas, the LLM is engaged in a loop of proposal, verification, and refinement. 

Similar to Algorithm~\ref{alg:non_agentic}, it starts with the same few-shot prompt (line~\ref{line:alg:agentic_prompt}). In each iteration, the LLM is asked to generate a set of candidate lemmas (line~\ref{alg:line:agentic_cand}). Upon receiving the candidate lemmas from the LLM, \lemmamine is invoked (line~\ref{alg:line:agentic_ind}) to determine whether the verification task has been \textit{solved}. That is, an inductive strengthening has been found. 
Throughout a conversation (i.e., the loop in lines~\ref{alg:line:loop_start}--\ref{alg:line:repair}), a record of all proposed lemmas is maintained in the set $All$. If the candidates generated by the LLM in a given iteration $i$ do not contain an inductive strengthening, \lemmamine is executed on the set of lemmas $All$ (line~\ref{alg:line:agentic_ind_all}). If an inductive strengthening is found in $All$, it is returned and the algorithm terminates.

Otherwise, a feedback message is constructed indicating that the generated candidates do not include an inductive strengthening (line~\ref{alg:line:repair}). The feedback also indicates for each suggested lemma whether it holds or not, based on the analysis performed by \lemmamine.
In addition, every two rounds, the feedback also includes general reminders of the task and the expected output format.



\subsection{Finding an Inductive Strengthening}\label{sec:find_ind}

Given the pair $(D,\varphi)$ and a set of candidate lemmas, 
we need to determine whether the candidate lemmas admit an inductive strengthening for the design and property. We implement an algorithm \lemmamine, which attempts to return a subset of lemmas that can be used as inductive strengthening, if such a subset exists. Since this algorithm is based on formal reasoning, if it determines that a subset is indeed an inductive strengthening, then we conclude that the property holds.

\begin{algorithm}[t]
\small
\caption{\lemmamine}\label{alg:lemma_mine}
\KwIn{A SystemVerilog design $D$, a property $\varphi$, and a set of candidate lemmas $\mathit{Cand}$}
\KwOut{An inductive strengthening $L$}
    $(T,AGp)\gets \texttt{GetTransitionSystemAndProp}(D,\varphi)$\;
    $C\gets\emptyset$, $Ind\gets\emptyset$, $All\gets\emptyset$\;
    \For {$\ell_i\in \mathit{Cand}$\label{alg:line:divide_begin}} {
        \If{\isoneind($T,\ell_i$)\label{alg:line:lemma-ind-check}} {
            $Ind\gets Ind\cup\{\ell_i\}$\;\label{alg:line:1-ind}
            \If{\isoneind($T,\ell_i\land p$)\label{alg:line:lemma-prop-ind-check}} {
                \Return {$\{\ell_i\}$}\;\label{alg:line:lemma-is-ind-s}
            }
        }
        \ElseIf{\propholds(T, $\ell_i, N$)} {
            $C\gets C\cup\{\ell_i\}$\;\label{alg:line:prop-holds}
        }
    }\label{alg:line:divide_end}
    $All\gets \text{\textsc{Concat}}(\text{\textsc{Sort}}(Ind), \text{\textsc{Sort}}(C))$\;\label{alg:line:sort}
    $L\gets\emptyset$\;
    \For {$i=1\ldots All.\text{\texttt{Size}}()$\label{alg:line:prefix-begin}} {
       \tcp{$L$ is the first $i$ elements of $All$}
       $L\gets L\cup\{All[i]\}$\;
       \If {\isoneind($T, \Land L\land p$)} {
            \Return {$L$}\;
       }
    }\label{alg:line:prefix-end}
    \Return {$\emptyset$}\;
\end{algorithm}


\lemmamine appears in Algorithm~\ref{alg:lemma_mine}.  To identify a
possible subset, \lemmamine partitions the set of candidates
into two sets of lemmas
(lines~\ref{alg:line:divide_begin}--\ref{alg:line:divide_end}): those that
are inductive invariants (line~\ref{alg:line:1-ind}), and those that hold up
to a given bound $N$ (line~\ref{alg:line:prop-holds}).  If one of the
candidate lemmas is
an inductive strengthening by itself
(line~\ref{alg:line:lemma-prop-ind-check}), the algorithm terminates and
returns it.  Candidate lemmas that fail both checks (e.g., do not hold and have a counterexample) are
discarded. 


Next, \lemmamine merges the two sets of inductive lemmas and lemmas that hold up to bound $N$ into a sorted list (line~\ref{alg:line:sort}). The list is ordered such that lemmas that are inductive invariants appear before lemmas that hold up to bound~$N$. Clearly, lemmas that are known to be inductive invariants have higher chances of being part of an inductive strengthening. For lemmas of the same category, a lexicographic order is used to enforce an ordering that is both deterministic and reproducible.

Lastly, \lemmamine iterates over the prefixes of the sorted list of lemmas and checks if some prefix is an inductive strengthening (lines~\ref{alg:line:prefix-begin}--\ref{alg:line:prefix-end}). If such a prefix is found, the algorithm terminates, and returns it. Otherwise, an empty set is returned, indicating that \lemmamine failed to find an inductive strengthening. Note that we only look for \emph{some} inductive strengthening, and not for a maximal or minimal subset (e.g., as in Houdini~\cite{DBLP:journals/ipl/FlanaganJL01})\footnote{There are many possible implementations that can be used to identify such a subset, e.g., Houdini, checking all possible subsets in parallel, or~\cite{6987603} and other similar approaches.}. Moreover, since \lemmamine does not consider all possible subsets, it may fail to find an inductive strengthening represented by a subset that is not a prefix. This limitation stems from the fact that not all lemmas in the sorted list are individually inductive.

\begin{algorithm}[t]
\small
\caption{\isoneind}\label{alg:isoneind}
\KwIn{A transition system $T$, and a formula $\psi$}
\KwOut{$\mathit{true}$ if $\psi$ is an inductive invariant w.r.t.~$T$, otherwise $\mathit{false}$}

    \If{$\textsc{IsSAT}(\Init(\Vars)\land\neg\psi(\Vars))$} {
        \Return $\mathit{false}$
    }
    \If{$\textsc{IsSAT}(\psi(\Vars)\land\Tr(\Vars,\Vars')\land\neg\psi(\Vars'))$} {
        \Return $\mathit{false}$
    }
    \Return $\mathit{true}$
\end{algorithm}

\lemmamine uses the function {\isoneind} to check if a property is an inductive invariant, and the function {\propholds} to check if the property has a counterexample of up to some given length.
The implementation of $\isoneind$ appears in Algorithm~\ref{alg:isoneind}. A call to $\isoneind(T,\psi)$ performs \emph{initiation} and
\emph{consecution} checks with respect to~the transition system $T$ and the property $\psi$ using a SAT-solver~\cite{DBLP:journals/pieee/VizelWM15} (see Definition~\ref{def:invar}). 
$\propholds(T,\psi,N)$ invokes a Bounded Model Checking
(BMC)~\cite{DBLP:journals/ac/BiereCCSZ03} engine with respect to the transition system $T$, a property $\psi$ and a bound $N$.

\begin{theorem}
If \lemmamine returns a non-empty set~$L$, then $L$ is an inductive strengthening for $\varphi$, and consequently $\varphi$ holds in the design $D$.
\end{theorem}

Following the above Theorem, we say that an example is \textit{solved} by an LLM if an invocation of \lemmamine on some set of candidate lemmas proposed by the LLM returns a non-empty set.

\section{Experimental Setup}\label{sec:experimental_setup}

\subsection{Prompt Configuration}
For the few-shot prompt (Algorithm~\ref{alg:non_agentic}), we experimented with $k=0,1,2,3,4$ few-shot examples and report the aggregate results across these configurations. We provide a per-$k$ breakdown in Table~\ref{tab:fewshot-ablation}.
For the Agentic setup, we allow a maximum of seven iterations ($n=7$ in Algorithm~\ref{alg:agentic}). 
While it is possible to provide counterexamples for suggested lemmas that fail verification, we empirically determined that 
it 
degraded the performance of the LLMs, possibly because including counterexamples made the feedback responses substantially longer.
The prompts we use can be found in~\cite{peled2026largelemmaminersllms,repo}.


\subsection{Formal Reasoning Configuration}
For the implementation of \lemmamine we used EBMC~\cite{mkm2015} (version 5.9). We check if a property is an inductive invariant using $k$-induction~\cite{DBLP:conf/fmcad/SheeranSS00} with $k=1$ (i.e., 1-inductive checks). For the bounded correctness checks (i.e., Bounded Model Checking~\cite{DBLP:journals/ac/BiereCCSZ03}), we use a bound of 30. We use bound 30 as a practical bound for bug-finding, sufficient to detect short counterexamples and quickly filter out incorrect candidates while also keeping the cost of this check manageable.
All queries use a timeout of 120 seconds. 

\subsection{LLMs and Inference Parameters}
We employ six
state-of-the-art LLMs: {\openai}'s \gptfive 
\footnote{snapshot \texttt{gpt-5-2025-08-07}} 
and {\anthropic}'s \claudefour, \claudefourfivesonnet, \claudefourfivehaiku, \claudefourfiveopus and \claudefoursevenopus \footnote{anthropic.claude-sonnet-4-20250514-v1:0,
anthropic.claude-sonnet-4-5-20250929-v1:0, anthropic.claude-haiku-4-5-20251001-v1:0, anthropic.claude-opus-4-5-20251101-v1:0 and
anthropic.claude-opus-4-7
in \texttt{Amazon Bedrock}, respectively}. 


For all Claude models, we take the default inference configuration provided with \texttt{Amazon Bedrock}'s \texttt{Converse API}. For \gptfive we take the default hyper-parameters assigned within {\openai}'s \texttt{Responses API}. 
We emphasize that {\openai}'s \texttt{Responses API} is used without supplying IDs of previous responses, and thus our prompting queries are {\bf stateless}.

\subsection{Re-Prompt Sampling}

In the case of the Agentic setup, we opt for sampling one response from the LLM for each sequence of messages.

For the Non-Agentic setup, we experiment with a varying number of samples, up to a total of five samples.
When sampling multiple responses in this setup, we need to decide how to invoke \lemmamine on them. In principle, for five sampled responses there are five sets of candidate lemmas. Namely, there are $2^5$ possible combinations we can consider. Each one of them gives rise to a different possible set of candidate lemmas that can be considered by \lemmamine. We opt for a simple approach in which we aggregate all candidate lemmas generated across all sampled responses into a single set and invoke \lemmamine on them all. 

\subsection{Dataset}\label{subsec:dataset}

    We compile a dataset of 57 \systemverilog designs with corresponding safety properties in SVA. Some modules have multiple associated properties, and some are parametrized and contribute multiple design instances. As discussed before, we consider one property per example, resulting in a total of 109 $(D, \varphi)$ pairs.  
    The designs and properties in our dataset are taken from several sources: \texttt{VCEGAR}~\cite{DBLP:conf/tacas/JainKSC07},  \texttt{NeuralMC}~\cite{neuralmc_25}, \texttt{VIS}~\cite{vis_benchmark},  \texttt{v2c}~\cite{DBLP:conf/tacas/MukherjeeTK16}, an Edge-Detector from~\cite{circuitcove}, and a counter from~\cite{interpolating_strong_induction}. In addition, we include one parametrized example from~\cite{vis_benchmark} (\texttt{Buffers}), and 3 additional parametrized designs of a FIFO (\texttt{Queues}), Least-Recently-Granted (\texttt{LRG}) arbiter and a content-addressed-memory (\texttt{CAM}).
    
    In addition to these hardware designs, we collected 47 short examples adapted from programs originally written in \texttt{C}. These examples originate from a benchmark suite for loop invariant generation~\cite{invgen_dataset,learning_invariants_using_llms}.
    Each example in this benchmark is a short \texttt{C} program containing a loop and a property to be verified (\textsf{assert} statement). To adapt these for our needs, we translate the \texttt{C} programs into simple finite state machines (FSMs) in \systemverilog. Of these examples, 16 are used to populate our \textit{CoT pool} described earlier. 
    For seven of these examples, we generate a total of 25 additional variants, which are examples with structurally similar underlying FSM, but small variations such as different parameter instantiations, different register sizes, swapped names for registers, a distracting extra register and/or an uninitialized register. Full details about the benchmarks are provided in~\cite{peled2026largelemmaminersllms}.

As our framework requires the RTL source code, the number of available open-source examples is limited. In particular, we cannot use numerous examples provided by the hardware model checking competitions (HWMCCs)~\cite{DBLP:conf/fmcad/BiereFP24}, as they are given in a \emph{netlist} format and do not include the original RTL code.

Lastly, 3 other examples from \cite{circuitcove} are used solely in the CoT pool. These examples are simple hardware constructs -- a frequency divider, a bidirectional counter and a 3-tap FIR, for which we manually write properties.


\section{Experimental Results}

The experiments were executed on Amazon EC2 
m6i.32xlarge instances with 128 vCPUs and 512\,GB memory for a total of 
1519.9 hours, of which $20\%$ were used by the LLMs and $80\%$ by the symbolic engine. Across all runs, the LLMs generated 
277.41 million tokens. 
The maximum time it took our framework to solve an example was 5 hours, with a mean and median of $232.54s$, and $85.1s$, respectively. Instructions for reproducing the results can be found in~\cite{repo}.

Results are summarized in Table~\ref{tab:overall}. As can be seen from the table, \claudefoursevenopus achieves the best performance. Overall, using all configurations, the Agentic and Non-Agentic setups solve 98 and 90 out of 109 examples, respectively. This accounts for a success rate of 90\% for the Agentic setup and 83\% for the Non-Agentic setup.
Excluding examples that are part of the CoT pool, or variants thereof, $84\%$ of the examples are solved by at least one of the configurations.



\begin{table*}[t]
  \centering
  \scalebox{0.8}{
  \begin{tabular}{@{}lrrrrrrrrrrr@{}}
     \toprule
    & \multicolumn{5}{c}{Agentic} & \multicolumn{5}{c}{Non-Agentic} \\
    \cmidrule(lr){2-6}\cmidrule(lr){7-11}
    Model & Total Lemmas & Correct & 1-Inductive & Error & Solved & Total Lemmas & Correct & 1-Inductive & Error & Solved \\
    \midrule
    \gptfive & 8090 & 5976 & 3187 & 175 & 84 & 12110 & 8647 & 5592 & 245 & 77 \\
    \claudefour & 6465 & 4651 & 2302 & 167 & 84 & 6720 & 4709 & 2839 & 307 & 68 \\
    \claudefourfivesonnet & 10255 & 7941 & 3636 & 203 & 89 & 8347 & 5982 & 3531 & 427 & 70 \\
    \claudefourfivehaiku & 10479 & 7189 & 3204 & 307 & 79 & 10384 & 6557 & 3521 & 503 & 71 \\
    \claudefourfiveopus & 8290 & 6851 & 3292 & 99 & 89 & 8498 & 6640 & 3956 & 271 & 84 \\
    \claudefoursevenopus & 6251 & 5025 & 2093 & 208 & 94 & 6233 & 4910 & 2840 & 284 & 88 \\
    \addlinespace
    Virtual Best &  &  &  &  & 98 &  &  &  &  & 90 \\
    \bottomrule
  \end{tabular}
             }
  \caption{Aggregate per (model, setup) across all main-experiment runs ($k\in\{0,1,2,3,4\}$ few-shot examples; up to five samples per Non-Agentic prompt). \textit{Total Lemmas}: candidate lemmas generated by the LLM. \textit{Error}: candidates the verifier could not parse or otherwise failed on. \textit{Correct}: candidates that hold in the design up to bound 30 (BMC). \textit{1-Inductive} ($\subseteq$ \textit{Correct}): 1-inductive invariants. \textit{Solved}: verification tasks (out of 109) for which \lemmamine extracted an inductive strengthening from the candidates. \textit{Virtual best}: tasks solved by at least one of the six models in the given setup.}
  \label{tab:overall}
\end{table*}

\subsection{SVA Familiarity}

We begin by noting that all LLMs occasionally exhibit gaps in their familiarity with SVA syntax. This further underscores the benefit of re-prompting to increase our collected pool of syntactically correct candidate lemmas.
\gptfive in one case incorrectly stated that {\bf \texttt{|->}} is SVA's non-overlapping operator, and provided a faulty reasoning step as a result. 
We have also observed an instance of an undefined variable $sum'$, referring to the next state of an existing register $sum$. 
Many models used invalid loop constructs in their assertions, particularly for lemmas involving indices of large registers.
We also encountered some encoding issues with {\claudefour}'s lemmas, where the generated lemmas contained non-ASCII characters, such as the Unicode symbol of the bidirectional arrow $\Leftrightarrow$ instead of the valid ASCII representation {\bf \texttt{<->}}. 
\claudefourfivehaiku has the highest error rate, roughly twice that of \gptfive.
Table~\ref{tab:overall} reports the number of errors per model and setup.



\subsection{Agentic vs.~Non-Agentic} 
We observe indicators for some relative advantage of the Agentic setup.
Note that the Agentic setup is in a sense more efficient: for half of the LLMs, the Agentic setup solves more examples than its Non-Agentic counterpart with a substantially smaller number of lemmas.
The apparent advantage of the Agentic setup can be attributed to the fact that \lemmamine is invoked on more subsets in this setup. 
Invoking \lemmamine on more subsets results in more queries to the verification tool with more combinations of candidate lemmas, which increases the chance of finding an inductive strengthening.
However, we have observed cases which suggest that the additional calls to \lemmamine are not the sole advantage of the Agentic setup over the Non-Agentic one. In these cases, a given LLM in the Non-Agentic setup fails to find a strengthening for a property, whereas its Agentic counterpart, over the course of several rounds, is steered away from incorrect or non-inductive lemmas and eventually proposes an inductive strengthening with a lemma that does not appear in the Non-Agentic setting at all. We therefore conjecture that the Agentic setup has an added value.

\subsection{Solved and Unsolved Examples}
Of 109 examples, 11 were not solved by any setting. A summary of 
the number of unsolved examples per category is provided in~\cite{peled2026largelemmaminersllms}. 
As expected, for all examples that were used to construct the CoT pool, the LLMs copied the lemmas they were shown, at times along with additional lemmas. Hence, all models solved the majority of this subset.
For the \texttt{C} programs group, the LLMs demonstrated no difficulty with swapped variable names and different parameter instantiations, but did struggle when the variants required an extra lemma accounting for possible overflow and/or accounting for an uninitialized variable.
Two unsolved examples from \texttt{v2c} are two of three examples
in our dataset that include properties with a sequence of three time frames ($\texttt{\#\#2}$) or more.
In fact, the design in one of these examples appears in another example in our dataset. However, the other example has a property that refers to two time frames ($\texttt{\#\#1}$), and it is solved. Another unsolved example from \texttt{v2c} 
is a conjunction of 6 conceptually distinct, operator-dense properties.

\subsection{Challenging Benchmarks}
The lemmas obtained in our experiments can vary widely in their complexity; for some designs, an inductive strengthening may be a simple auxiliary invariant, whereas others may require nontrivial ones. In this subsection, we therefore 
 target a more stringent question---Can LLMs derive \textit{nontrivial} lemmas when constructing an inductive strengthening?
 It is hard to characterize and agree upon what constitutes a ``nontrivial'' lemma. Therefore, we adopt a pragmatic approach wherein we use SOTA model checkers as a proxy for nontriviality. 
 Notable SOTA tools in our domain are Cadence's JasperGold (version 2024.06p002),
Synopsys' VC-Formal (version X-2025.06-SP2P3)
 and the open-source IC3-based rIC3~\cite{DBLP:conf/cav/SuYCBH25} (version 1.5.2).
 We consider an example to require a \textit{nontrivial} lemma if one of the aforementioned tools does not prove its property in under 10 minutes. We justify this decision by the fact that the SOTA tools in our domain are based on mature and well-established algorithms, backed by decades of research and extensive industrial adoption.


We identify 
31 such challenging examples in our dataset:

\begin{enumerate*}[label=(\roman*)]
    \item a content-addressed memory design (\texttt{CAM}, 2 instances of different sizes); 
    \item a Least-Recently-Granted arbiter (\texttt{LRG}, 3 instances);
    \item a buffer design from \texttt{vis} (3 instances);
    \item two designs checking equivalence between FIFOs (\texttt{Queues}). One of these designs is parametrized and contributes two instances. Others are drawn from \texttt{VIS} and the additional categories of our dataset;  
    \item 20 counter designs (\texttt{Counters}), 18 drawn from the \texttt{C} programs category, one from \texttt{VCEGAR}, and one example adapted from \cite{interpolating_strong_induction}.
\end{enumerate*}

The comparison of our approach and the aforementioned SOTA tools on these examples is summarized in Table~\ref{tab:hard_benchmarks}.
A~timeout of one hour was used for all tools.


Our pipeline solved 27 out of these 31 examples.
For the larger buffer, the LLMs produced
a valid inductive strengthening, but the subsequent induction check exceeded our given timeout of 2 minutes. We note that JasperGold can complete that induction check within the 2 minute timeout and report  this example as 
TIMEOUT$^\dagger$. Another example was solved by the Agentic \claudefoursevenopus in 2 hours which exceeds the given timeout of one hour, and we report it as TIMEOUT$^\ddagger$. 
Interestingly, for the larger LRG-arbiter instances, the main obstacle for some LLMs such as \claudefourfivesonnet
was using invalid loop constructs in their lemmas.  Enumerating the loops explicitly 
does yield valid inductive strengthenings.
\begin{table}[t]
\resizebox{0.6\textwidth}{!}{\begin{minipage}{\textwidth}
\begin{tabular}{llrrrr}
\toprule
Category & Design & Our virtual best & rIC3 & JasperGold & VCF \\
\midrule
\multirow{2}{*}{CAM} & simple\_cam\_16 & 356.06 & 0.06 & 188.00 & TIMEOUT \\
 & simple\_cam\_8 & 22.98 & 0.03 & 187.00 & TIMEOUT \\
\cmidrule{1-6}
\multirow{3}{*}{arbiters} & lrg\_arb\_lrg\_16 & TIMEOUT & TIMEOUT & 907.00 & 750.00 \\
 & lrg\_arb\_lrg\_4 & 28.09 & 0.02 & 3.00 & 1.00 \\
 & lrg\_arb\_lrg\_8 & TIMEOUT & 2.55 & 5.00 & 2.00 \\
\cmidrule{1-6}
\multirow{3}{*}{buffers} & buffer\_128 & TIMEOUT$^\dagger$ & 0.06 & TIMEOUT & TIMEOUT \\
 & buffer\_32 & 261.00 & 0.01 & 188.00 & TIMEOUT \\
 & buffer\_64 & 419.10 & 0.03 & 435.00 & TIMEOUT \\
\cmidrule{1-6}
\multirow{20}{*}{counters} & counter\_66 & 12.27 & TIMEOUT & TIMEOUT & 15.00 \\
 & dp & 15.10 & TIMEOUT & 6.00 & TIMEOUT \\
 & ex15 & 9.91 & TIMEOUT & 187.00 & TIMEOUT \\
 & ex3 & 7.08 & TIMEOUT & 54.00 & 1.00 \\
 & ex4 & 6.89 & TIMEOUT & 3.00 & 1.00 \\
 & ex51 & 8.01 & TIMEOUT & 4.00 & 2.00 \\
 & ex51\_57\_1 & 19.32 & TIMEOUT & 6.00 & TIMEOUT \\
 & ex80 & 13.10 & TIMEOUT & 2340.00 & TIMEOUT \\
 & gr2006 & 10.42 & 275.51 & 820.00 & 1.00 \\
 & gulwani\_cegar2 & 8.21 & TIMEOUT & 187.00 & TIMEOUT \\
 & gulwani\_cegar2\_1 & 18.79 & 123.82 & 287.00 & TIMEOUT \\
 & gulwani\_cegar2\_2 & 19.60 & 119.36 & 356.00 & TIMEOUT \\
 & gulwani\_cegar2\_3 & 25.26 & TIMEOUT & 233.00 & TIMEOUT \\
 & gulwani\_cegar2\_4 & 21.27 & TIMEOUT & 218.00 & TIMEOUT \\
 & gulwani\_cegar2\_5 & 29.40 & TIMEOUT & TIMEOUT & TIMEOUT \\
 & gulwani\_fig1a & 731.92 & TIMEOUT & 192.00 & 1.00 \\
 & gulwani\_fig1a\_2 & TIMEOUT$^\ddagger$ & TIMEOUT & 853.00 & 1.00 \\
 & gulwani\_fig1a\_4 & 859.28 & TIMEOUT & 613.00 & 1.00 \\
 & gulwani\_fig1a\_5 & 1235.65 & TIMEOUT & 192.00 & 1.00 \\
 & gulwani\_fig1a\_6 & 970.68 & TIMEOUT & TIMEOUT & 1.00 \\
\cmidrule{1-6}
\multirow{3}{*}{queues} & fifo\_ref\_16 & 96.93 & 0.04 & TIMEOUT & TIMEOUT \\
 & fifo\_ref\_64 & 195.33 & 0.15 & TIMEOUT & TIMEOUT \\
 & fifo\_vis & 849.16 & 619.63 & 1580.00 & TIMEOUT \\
\bottomrule
\end{tabular}
\end{minipage} 
}
\caption{Runtime comparison (in seconds)}
\label{tab:hard_benchmarks}
\end{table}

\subsection{Ablations}\label{sec:ablations}

\subsubsection{Number of CoT Examples}
The Few-Shot prompt can be constructed using an arbitrary number $k$ of examples drawn from the CoT pool. In our experiments, we considered $k \in \{ 0,1,2,3,4\}$ and reported results aggregated across these five choices.  To assess the sensitivity of our framework to the parameter $k$ (i.e. the number of CoT examples used in the prompt), we evaluate our framework with a fixed $0 \leq k\leq 5$.
Table ~\ref{tab:fewshot-ablation} summarizes the results of this experiment. For each configuration, the table presents the percentage of examples solved for every value of $0\leq k\leq 5$. Darker shades of blue indicate a higher success rate. For every LLM, the results are broken down by three dataset categories:
\begin{enumerate*}
    \item \textit{seen} examples, i.e., examples that appear in the CoT pool;
    \item \textit{C examples} that do not appear in the CoT pool (i.e., from \textit{c, generalize} and \textit{c, others});
    \item and all remaining designs (from \texttt{VCEGAR}, \texttt{VIS}, \texttt{v2c}, \texttt{NeuralMC}, and the additional parametrized designs).
\end{enumerate*}
The \textsc{VB} (Virtual Best) column aggregates the results across the different values of $k$, while the Virtual Best row presents the aggregation across all LLMs.

\definecolor{pctred1}{HTML}{F4CCCC}
\definecolor{pctred2}{HTML}{FCD5B4}
\definecolor{pctmid1}{HTML}{FFE599}
\definecolor{pctmid2}{HTML}{FFF2CC}
\definecolor{pctgrn1}{HTML}{E2EFDA}
\definecolor{pctgrn2}{HTML}{C6E0B4}
\definecolor{pctgrn3}{HTML}{A9D08E}
\begin{table*}
\centering
\resizebox{0.85\textwidth}{!}{%
\begin{tabular}{lllrrrrrrr||rrrrrrr}
\toprule
\textbf{Model} & \textbf{Tag} & \textbf{Total} & \multicolumn{7}{c}{Agentic} & \multicolumn{7}{c}{Non-Agentic} \\
 &  &  & $k=0$ & $k=1$ & $k=2$ & $k=3$ & $k=4$ & $k=5$ & $\textsc{VB}$ & $k=0$ & $k=1$ & $k=2$ & $k=3$ & $k=4$ & $k=5$ & $\textsc{VB}$ \\
\midrule
\multirow{3}{*}{\claudefour} & c examples & 31 & \cellcolor[RGB]{35,115,182}\color{white}74\% & \cellcolor[RGB]{15,90,163}\color{white}\textbf{84\%} & \cellcolor[RGB]{44,124,186}\color{white}71\% & \cellcolor[RGB]{22,99,170}\color{white}81\% & \cellcolor[RGB]{28,107,176}\color{white}77\% & \cellcolor[RGB]{22,99,170}\color{white}81\% & \cellcolor[RGB]{8,65,132}\color{white}94\% & \cellcolor[RGB]{52,132,191}\color{white}\textbf{68\%} & \cellcolor[RGB]{61,141,195}\color{white}65\% & \cellcolor[RGB]{70,149,200}\color{white}61\% & \cellcolor[RGB]{81,156,204}58\% & \cellcolor[RGB]{70,149,200}\color{white}61\% & \cellcolor[RGB]{61,141,195}\color{white}65\% & \cellcolor[RGB]{35,115,182}\color{white}74\% \\
 & others & 60 & \cellcolor[RGB]{107,174,214}50\% & \cellcolor[RGB]{114,178,215}48\% & \cellcolor[RGB]{114,178,215}48\% & \cellcolor[RGB]{121,181,217}47\% & \cellcolor[RGB]{102,170,212}\textbf{52\%} & \cellcolor[RGB]{114,178,215}48\% & \cellcolor[RGB]{69,148,199}\color{white}62\% & \cellcolor[RGB]{148,196,223}\textbf{40\%} & \cellcolor[RGB]{155,200,224}38\% & \cellcolor[RGB]{161,203,226}37\% & \cellcolor[RGB]{161,203,226}37\% & \cellcolor[RGB]{155,200,224}38\% & \cellcolor[RGB]{155,200,224}38\% & \cellcolor[RGB]{121,181,217}47\% \\
 & seen & 18 & \cellcolor[RGB]{40,120,185}\color{white}72\% & \cellcolor[RGB]{8,63,129}\color{white}94\% & \cellcolor[RGB]{8,63,129}\color{white}94\% & \cellcolor[RGB]{8,63,129}\color{white}94\% & \cellcolor[RGB]{8,48,107}\color{white}\textbf{100\%} & \cellcolor[RGB]{8,77,151}\color{white}89\% & \cellcolor[RGB]{8,48,107}\color{white}100\% & \cellcolor[RGB]{71,149,200}\color{white}61\% & \cellcolor[RGB]{8,63,129}\color{white}\textbf{94\%} & \cellcolor[RGB]{8,63,129}\color{white}\textbf{94\%} & \cellcolor[RGB]{8,77,151}\color{white}89\% & \cellcolor[RGB]{8,63,129}\color{white}\textbf{94\%} & \cellcolor[RGB]{8,77,151}\color{white}89\% & \cellcolor[RGB]{8,63,129}\color{white}94\% \\
\midrule
\multirow{3}{*}{\claudefourfivehaiku} & c examples & 31 & \cellcolor[RGB]{70,149,200}\color{white}61\% & \cellcolor[RGB]{44,124,186}\color{white}71\% & \cellcolor[RGB]{52,132,191}\color{white}68\% & \cellcolor[RGB]{35,115,182}\color{white}\textbf{74\%} & \cellcolor[RGB]{35,115,182}\color{white}\textbf{74\%} & \cellcolor[RGB]{61,141,195}\color{white}65\% & \cellcolor[RGB]{28,107,176}\color{white}77\% & \cellcolor[RGB]{102,170,212}52\% & \cellcolor[RGB]{81,156,204}58\% & \cellcolor[RGB]{70,149,200}\color{white}\textbf{61\%} & \cellcolor[RGB]{81,156,204}58\% & \cellcolor[RGB]{70,149,200}\color{white}\textbf{61\%} & \cellcolor[RGB]{70,149,200}\color{white}\textbf{61\%} & \cellcolor[RGB]{15,90,163}\color{white}84\% \\
 & others & 60 & \cellcolor[RGB]{148,196,223}40\% & \cellcolor[RGB]{166,205,228}35\% & \cellcolor[RGB]{166,205,228}35\% & \cellcolor[RGB]{121,181,217}\textbf{47\%} & \cellcolor[RGB]{127,185,218}45\% & \cellcolor[RGB]{141,193,221}42\% & \cellcolor[RGB]{55,135,192}\color{white}67\% & \cellcolor[RGB]{166,205,228}35\% & \cellcolor[RGB]{166,205,228}35\% & \cellcolor[RGB]{148,196,223}\textbf{40\%} & \cellcolor[RGB]{161,203,226}37\% & \cellcolor[RGB]{161,203,226}37\% & \cellcolor[RGB]{177,210,232}32\% & \cellcolor[RGB]{107,174,214}50\% \\
 & seen & 18 & \cellcolor[RGB]{107,174,214}50\% & \cellcolor[RGB]{8,77,151}\color{white}\textbf{89\%} & \cellcolor[RGB]{40,120,185}\color{white}72\% & \cellcolor[RGB]{8,77,151}\color{white}\textbf{89\%} & \cellcolor[RGB]{27,106,175}\color{white}78\% & \cellcolor[RGB]{8,77,151}\color{white}\textbf{89\%} & \cellcolor[RGB]{8,63,129}\color{white}94\% & \cellcolor[RGB]{89,162,207}56\% & \cellcolor[RGB]{8,63,129}\color{white}\textbf{94\%} & \cellcolor[RGB]{8,63,129}\color{white}\textbf{94\%} & \cellcolor[RGB]{8,63,129}\color{white}\textbf{94\%} & \cellcolor[RGB]{8,77,151}\color{white}89\% & \cellcolor[RGB]{8,77,151}\color{white}89\% & \cellcolor[RGB]{8,63,129}\color{white}94\% \\
\midrule
\multirow{3}{*}{\claudefourfiveopus} & c examples & 31 & \cellcolor[RGB]{35,115,182}\color{white}74\% & \cellcolor[RGB]{22,99,170}\color{white}81\% & \cellcolor[RGB]{28,107,176}\color{white}77\% & \cellcolor[RGB]{15,90,163}\color{white}\textbf{84\%} & \cellcolor[RGB]{28,107,176}\color{white}77\% & \cellcolor[RGB]{22,99,170}\color{white}81\% & \cellcolor[RGB]{8,74,145}\color{white}90\% & \cellcolor[RGB]{28,107,176}\color{white}77\% & \cellcolor[RGB]{15,90,163}\color{white}\textbf{84\%} & \cellcolor[RGB]{35,115,182}\color{white}74\% & \cellcolor[RGB]{52,132,191}\color{white}68\% & \cellcolor[RGB]{52,132,191}\color{white}68\% & \cellcolor[RGB]{44,124,186}\color{white}71\% & \cellcolor[RGB]{9,82,157}\color{white}87\% \\
 & others & 60 & \cellcolor[RGB]{51,131,190}\color{white}68\% & \cellcolor[RGB]{42,122,186}\color{white}\textbf{72\%} & \cellcolor[RGB]{55,135,192}\color{white}67\% & \cellcolor[RGB]{64,144,197}\color{white}63\% & \cellcolor[RGB]{55,135,192}\color{white}67\% & \cellcolor[RGB]{46,126,188}\color{white}70\% & \cellcolor[RGB]{37,117,183}\color{white}73\% & \cellcolor[RGB]{74,152,201}\textbf{60\%} & \cellcolor[RGB]{74,152,201}\textbf{60\%} & \cellcolor[RGB]{85,159,205}57\% & \cellcolor[RGB]{74,152,201}\textbf{60\%} & \cellcolor[RGB]{85,159,205}57\% & \cellcolor[RGB]{74,152,201}\textbf{60\%} & \cellcolor[RGB]{51,131,190}\color{white}68\% \\
 & seen & 18 & \cellcolor[RGB]{8,77,151}\color{white}89\% & \cellcolor[RGB]{8,48,107}\color{white}\textbf{100\%} & \cellcolor[RGB]{8,63,129}\color{white}94\% & \cellcolor[RGB]{8,63,129}\color{white}94\% & \cellcolor[RGB]{8,63,129}\color{white}94\% & \cellcolor[RGB]{8,63,129}\color{white}94\% & \cellcolor[RGB]{8,48,107}\color{white}100\% & \cellcolor[RGB]{16,92,164}\color{white}83\% & \cellcolor[RGB]{8,63,129}\color{white}\textbf{94\%} & \cellcolor[RGB]{8,63,129}\color{white}\textbf{94\%} & \cellcolor[RGB]{8,77,151}\color{white}89\% & \cellcolor[RGB]{8,63,129}\color{white}\textbf{94\%} & \cellcolor[RGB]{8,77,151}\color{white}89\% & \cellcolor[RGB]{8,63,129}\color{white}94\% \\
\midrule
\multirow{3}{*}{\claudefourfivesonnet} & c examples & 31 & \cellcolor[RGB]{22,99,170}\color{white}81\% & \cellcolor[RGB]{9,82,157}\color{white}\textbf{87\%} & \cellcolor[RGB]{22,99,170}\color{white}81\% & \cellcolor[RGB]{15,90,163}\color{white}84\% & \cellcolor[RGB]{9,82,157}\color{white}\textbf{87\%} & \cellcolor[RGB]{9,82,157}\color{white}\textbf{87\%} & \cellcolor[RGB]{8,57,120}\color{white}97\% & \cellcolor[RGB]{70,149,200}\color{white}61\% & \cellcolor[RGB]{61,141,195}\color{white}65\% & \cellcolor[RGB]{70,149,200}\color{white}61\% & \cellcolor[RGB]{44,124,186}\color{white}\textbf{71\%} & \cellcolor[RGB]{70,149,200}\color{white}61\% & \cellcolor[RGB]{81,156,204}58\% & \cellcolor[RGB]{44,124,186}\color{white}71\% \\
 & others & 60 & \cellcolor[RGB]{74,152,201}60\% & \cellcolor[RGB]{80,155,203}58\% & \cellcolor[RGB]{80,155,203}58\% & \cellcolor[RGB]{102,170,212}52\% & \cellcolor[RGB]{80,155,203}58\% & \cellcolor[RGB]{64,144,197}\color{white}\textbf{63\%} & \cellcolor[RGB]{46,126,188}\color{white}70\% & \cellcolor[RGB]{127,185,218}45\% & \cellcolor[RGB]{134,189,220}43\% & \cellcolor[RGB]{134,189,220}43\% & \cellcolor[RGB]{155,200,224}38\% & \cellcolor[RGB]{166,205,228}35\% & \cellcolor[RGB]{114,178,215}\textbf{48\%} & \cellcolor[RGB]{85,159,205}57\% \\
 & seen & 18 & \cellcolor[RGB]{8,77,151}\color{white}89\% & \cellcolor[RGB]{8,63,129}\color{white}94\% & \cellcolor[RGB]{8,63,129}\color{white}94\% & \cellcolor[RGB]{8,63,129}\color{white}94\% & \cellcolor[RGB]{8,48,107}\color{white}\textbf{100\%} & \cellcolor[RGB]{8,48,107}\color{white}\textbf{100\%} & \cellcolor[RGB]{8,48,107}\color{white}100\% & \cellcolor[RGB]{40,120,185}\color{white}72\% & \cellcolor[RGB]{8,63,129}\color{white}\textbf{94\%} & \cellcolor[RGB]{8,77,151}\color{white}89\% & \cellcolor[RGB]{8,77,151}\color{white}89\% & \cellcolor[RGB]{8,77,151}\color{white}89\% & \cellcolor[RGB]{8,77,151}\color{white}89\% & \cellcolor[RGB]{8,63,129}\color{white}94\% \\
\midrule
\multirow{3}{*}{\claudefoursevenopus} & c examples & 31 & \cellcolor[RGB]{8,74,145}\color{white}90\% & \cellcolor[RGB]{8,74,145}\color{white}90\% & \cellcolor[RGB]{8,65,132}\color{white}94\% & \cellcolor[RGB]{8,74,145}\color{white}90\% & \cellcolor[RGB]{8,74,145}\color{white}90\% & \cellcolor[RGB]{8,48,107}\color{white}\textbf{100\%} & \cellcolor[RGB]{8,48,107}\color{white}100\% & \cellcolor[RGB]{28,107,176}\color{white}77\% & \cellcolor[RGB]{15,90,163}\color{white}84\% & \cellcolor[RGB]{22,99,170}\color{white}81\% & \cellcolor[RGB]{9,82,157}\color{white}\textbf{87\%} & \cellcolor[RGB]{44,124,186}\color{white}71\% & \cellcolor[RGB]{35,115,182}\color{white}74\% & \cellcolor[RGB]{8,57,120}\color{white}97\% \\
 & others & 60 & \cellcolor[RGB]{55,135,192}\color{white}67\% & \cellcolor[RGB]{37,117,183}\color{white}\textbf{73\%} & \cellcolor[RGB]{46,126,188}\color{white}70\% & \cellcolor[RGB]{46,126,188}\color{white}70\% & \cellcolor[RGB]{37,117,183}\color{white}\textbf{73\%} & \cellcolor[RGB]{42,122,186}\color{white}72\% & \cellcolor[RGB]{33,113,181}\color{white}75\% & \cellcolor[RGB]{74,152,201}60\% & \cellcolor[RGB]{74,152,201}60\% & \cellcolor[RGB]{69,148,199}\color{white}62\% & \cellcolor[RGB]{74,152,201}60\% & \cellcolor[RGB]{64,144,197}\color{white}\textbf{63\%} & \cellcolor[RGB]{85,159,205}57\% & \cellcolor[RGB]{51,131,190}\color{white}68\% \\
 & seen & 18 & \cellcolor[RGB]{8,48,107}\color{white}\textbf{100\%} & \cellcolor[RGB]{8,48,107}\color{white}\textbf{100\%} & \cellcolor[RGB]{8,48,107}\color{white}\textbf{100\%} & \cellcolor[RGB]{8,48,107}\color{white}\textbf{100\%} & \cellcolor[RGB]{8,48,107}\color{white}\textbf{100\%} & \cellcolor[RGB]{8,48,107}\color{white}\textbf{100\%} & \cellcolor[RGB]{8,48,107}\color{white}100\% & \cellcolor[RGB]{8,77,151}\color{white}89\% & \cellcolor[RGB]{8,63,129}\color{white}94\% & \cellcolor[RGB]{8,63,129}\color{white}94\% & \cellcolor[RGB]{8,48,107}\color{white}\textbf{100\%} & \cellcolor[RGB]{8,63,129}\color{white}94\% & \cellcolor[RGB]{8,77,151}\color{white}89\% & \cellcolor[RGB]{8,48,107}\color{white}100\% \\
\midrule
\multirow{3}{*}{\gptfive} & c examples & 31 & \cellcolor[RGB]{35,115,182}\color{white}74\% & \cellcolor[RGB]{28,107,176}\color{white}77\% & \cellcolor[RGB]{44,124,186}\color{white}71\% & \cellcolor[RGB]{28,107,176}\color{white}77\% & \cellcolor[RGB]{35,115,182}\color{white}74\% & \cellcolor[RGB]{22,99,170}\color{white}\textbf{81\%} & \cellcolor[RGB]{9,82,157}\color{white}87\% & \cellcolor[RGB]{35,115,182}\color{white}74\% & \cellcolor[RGB]{28,107,176}\color{white}77\% & \cellcolor[RGB]{22,99,170}\color{white}\textbf{81\%} & \cellcolor[RGB]{22,99,170}\color{white}\textbf{81\%} & \cellcolor[RGB]{22,99,170}\color{white}\textbf{81\%} & \cellcolor[RGB]{28,107,176}\color{white}77\% & \cellcolor[RGB]{15,90,163}\color{white}84\% \\
 & others & 60 & \cellcolor[RGB]{107,174,214}50\% & \cellcolor[RGB]{107,174,214}50\% & \cellcolor[RGB]{102,170,212}52\% & \cellcolor[RGB]{91,163,208}\textbf{55\%} & \cellcolor[RGB]{114,178,215}48\% & \cellcolor[RGB]{114,178,215}48\% & \cellcolor[RGB]{46,126,188}\color{white}70\% & \cellcolor[RGB]{114,178,215}\textbf{48\%} & \cellcolor[RGB]{134,189,220}43\% & \cellcolor[RGB]{148,196,223}40\% & \cellcolor[RGB]{134,189,220}43\% & \cellcolor[RGB]{127,185,218}45\% & \cellcolor[RGB]{141,193,221}42\% & \cellcolor[RGB]{85,159,205}57\% \\
 & seen & 18 & \cellcolor[RGB]{40,120,185}\color{white}72\% & \cellcolor[RGB]{8,63,129}\color{white}\textbf{94\%} & \cellcolor[RGB]{8,63,129}\color{white}\textbf{94\%} & \cellcolor[RGB]{8,63,129}\color{white}\textbf{94\%} & \cellcolor[RGB]{8,63,129}\color{white}\textbf{94\%} & \cellcolor[RGB]{8,63,129}\color{white}\textbf{94\%} & \cellcolor[RGB]{8,63,129}\color{white}94\% & \cellcolor[RGB]{8,77,151}\color{white}89\% & \cellcolor[RGB]{8,77,151}\color{white}89\% & \cellcolor[RGB]{8,63,129}\color{white}\textbf{94\%} & \cellcolor[RGB]{8,63,129}\color{white}\textbf{94\%} & \cellcolor[RGB]{8,77,151}\color{white}89\% & \cellcolor[RGB]{8,63,129}\color{white}\textbf{94\%} & \cellcolor[RGB]{8,63,129}\color{white}94\% \\
\midrule
\multirow{3}{*}{\textbf{Virtual best}} & c examples & 31 & \cellcolor[RGB]{8,74,145}\color{white}90\% & \cellcolor[RGB]{8,65,132}\color{white}94\% & \cellcolor[RGB]{8,65,132}\color{white}94\% & \cellcolor[RGB]{8,48,107}\color{white}100\% & \cellcolor[RGB]{8,57,120}\color{white}97\% & \cellcolor[RGB]{8,48,107}\color{white}100\% & \cellcolor[RGB]{8,48,107}\color{white}100\% & \cellcolor[RGB]{15,90,163}\color{white}84\% & \cellcolor[RGB]{9,82,157}\color{white}87\% & \cellcolor[RGB]{9,82,157}\color{white}87\% & \cellcolor[RGB]{8,65,132}\color{white}94\% & \cellcolor[RGB]{15,90,163}\color{white}84\% & \cellcolor[RGB]{9,82,157}\color{white}87\% & \cellcolor[RGB]{8,57,120}\color{white}97\% \\
 & others & 60 & \cellcolor[RGB]{37,117,183}\color{white}73\% & \cellcolor[RGB]{33,113,181}\color{white}75\% & \cellcolor[RGB]{42,122,186}\color{white}72\% & \cellcolor[RGB]{42,122,186}\color{white}72\% & \cellcolor[RGB]{30,109,178}\color{white}77\% & \cellcolor[RGB]{33,113,181}\color{white}75\% & \cellcolor[RGB]{20,96,168}\color{white}82\% & \cellcolor[RGB]{51,131,190}\color{white}68\% & \cellcolor[RGB]{55,135,192}\color{white}67\% & \cellcolor[RGB]{55,135,192}\color{white}67\% & \cellcolor[RGB]{59,139,195}\color{white}65\% & \cellcolor[RGB]{55,135,192}\color{white}67\% & \cellcolor[RGB]{55,135,192}\color{white}67\% & \cellcolor[RGB]{42,122,186}\color{white}72\% \\
 & seen & 18 & \cellcolor[RGB]{8,48,107}\color{white}100\% & \cellcolor[RGB]{8,48,107}\color{white}100\% & \cellcolor[RGB]{8,48,107}\color{white}100\% & \cellcolor[RGB]{8,48,107}\color{white}100\% & \cellcolor[RGB]{8,48,107}\color{white}100\% & \cellcolor[RGB]{8,48,107}\color{white}100\% & \cellcolor[RGB]{8,48,107}\color{white}100\% & \cellcolor[RGB]{8,77,151}\color{white}89\% & \cellcolor[RGB]{8,63,129}\color{white}94\% & \cellcolor[RGB]{8,63,129}\color{white}94\% & \cellcolor[RGB]{8,48,107}\color{white}100\% & \cellcolor[RGB]{8,63,129}\color{white}94\% & \cellcolor[RGB]{8,63,129}\color{white}94\% & \cellcolor[RGB]{8,48,107}\color{white}100\% \\
\bottomrule
\end{tabular}
}
\caption{Few-shot ablation results}
\label{tab:fewshot-ablation}
\end{table*}


We observe some variation in the sensitivity of different settings to the number of CoT examples. The Agentic \claudefoursevenopus configuration is relatively robust, while Agentic \claudefourfivehaiku is more affected.
Increasing the number of few-shot examples does not produce a monotonic improvement, but omitting the few-shot examples altogether does reduce performance: the virtual-best success rate increases from $82\%$ of examples for $k=0$ to $86\%$ for $k=4$. We do not observe an improvement when increasing the number of few-shot examples to $k=5$.


\subsubsection{Number of Iterations for the LLM Agent}

\begin{figure}[t]
    \centering
    \includegraphics[width=0.98\linewidth]{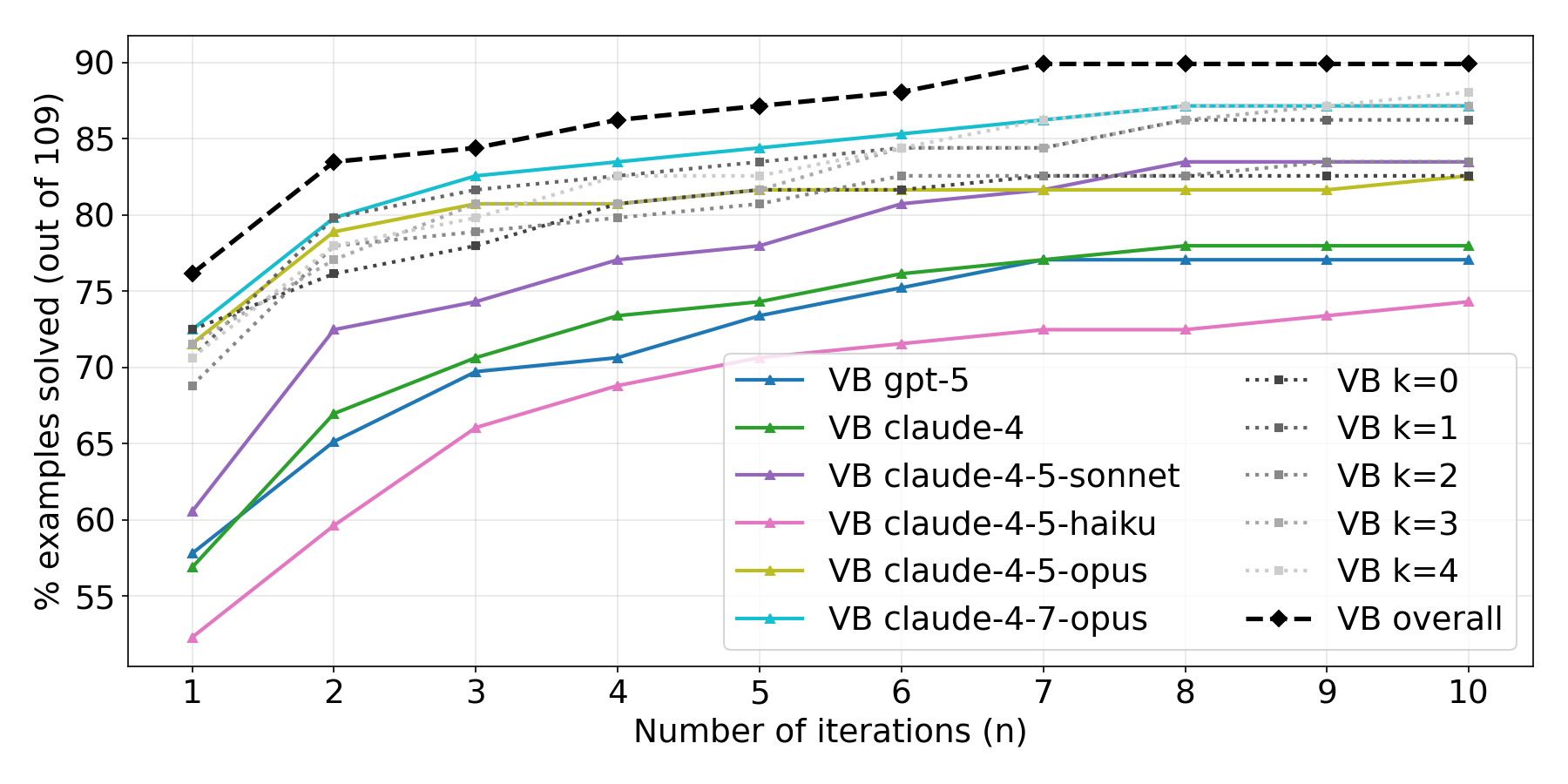}
    \centering 
    \caption{Examples solved by number of Agentic iterations. "VB"  (virtual best) per-model: examples solved by at least one few-shot configuration for that model; "VB" per few-shot: examples solved by at least one of the models for that few-shot configuration; "VB" overall: examples solved by at least one of the models and few-shot configurations.}
    \label{fig:iterations_ablation}
\end{figure}

\begin{figure}[t]
    \centering
    \includegraphics[width=0.85\linewidth]{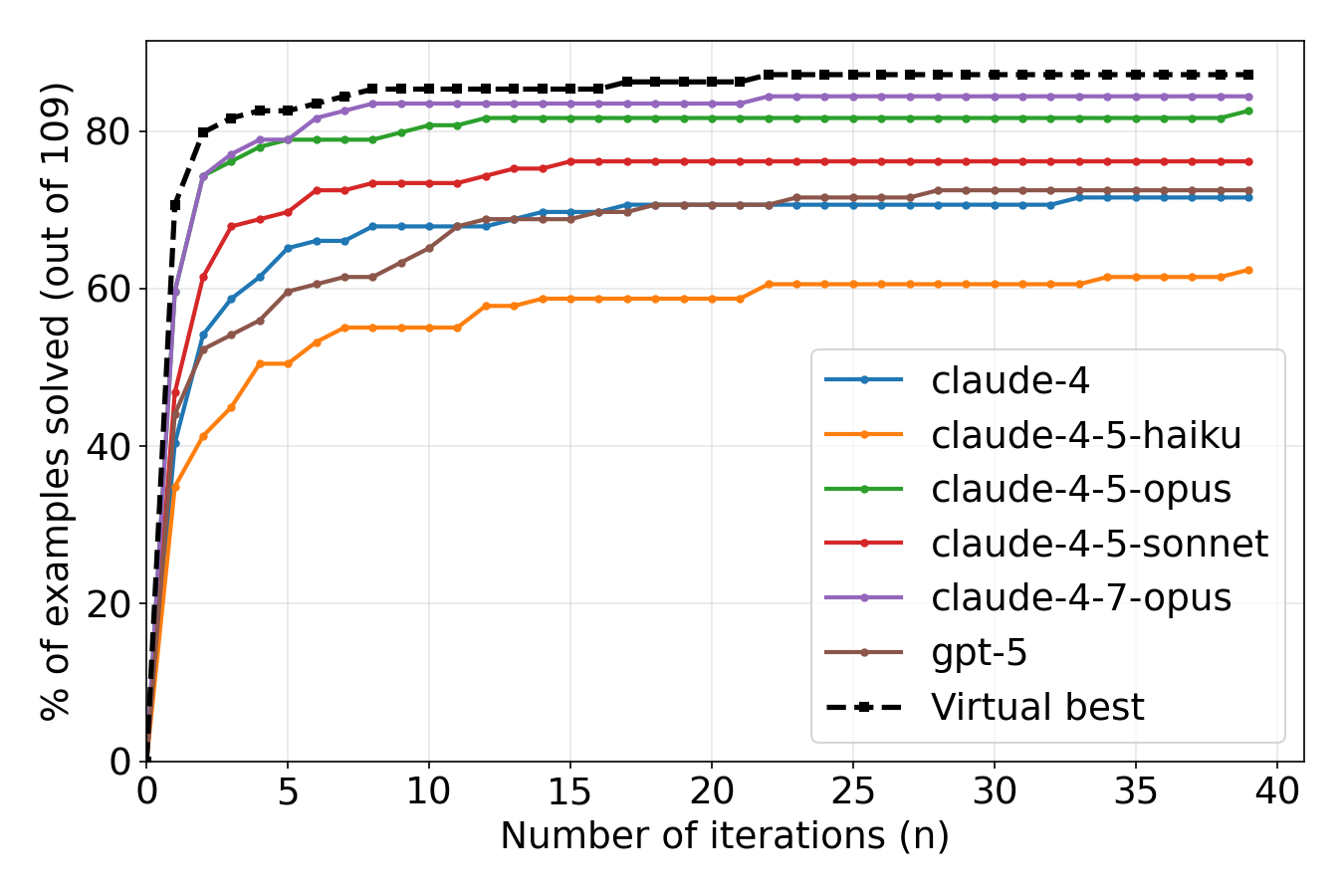}
    \centering 
    \caption{Examples solved within 2 hours by number of Agentic iterations. Results are shown for configurations with $k=1$ few-shot examples.}
    \label{fig:iterations_cactus}
\end{figure}

In the Agentic setup, the LLM interacts with the symbolic engine for $n$ iterations. To assess how sensitive our framework is to the iteration count $n$, we evaluate the Agentic framework across $1 \leq n \leq 10$ iterations, and report the results in Figure~\ref{fig:iterations_ablation}. For each $n$, we provide the virtual best (VB) results broken down by model, by parameter $0 \leq k \leq 4$, and aggregated across all models and $k$ values.
In addition, we perform the following experiment:
We fix the number of few-shot examples to $k=1$, bound the Agentic setup with a timeout of 2 hours and let it use as many iterations as it needs (no bound on $n$) to complete the verification task. Figure~\ref{fig:iterations_cactus} presents a cactus plot that summarizes the results of this experiment. The X-axis represents the number of iterations $n$, while the Y-axis represents the percentage of solved examples. As can be seen from the plot, the majority of examples are solved with up to 5 iterations ($n\leq 5)$, while all LLMs saturate within approximately 10 iterations.

\section{Limitations and Threats to Validity}

\subsection{Data Contamination}
Data contamination is a well-known concern when evaluating LLMs on generalization tasks. 
If benchmark instances or fragments thereof appear in pretraining data, strong performance may result from memorization, and not from genuine reasoning capabilities of the LLM, undermining the validity of the evaluation. 
Some of the examples in our dataset have been publicly available long enough to plausibly appear in the pretraining corpus of the LLMs we evaluate. However, to the best of our knowledge, there are no available auxiliary lemmas which could serve as inductive strengthenings for them.
More specifically, the benchmarks collected from \cite{vis_benchmark,DBLP:conf/tacas/MukherjeeTK16,DBLP:conf/tacas/JainKSC07} were originally designed for a different task, which reduces the risk of reference answers being available. 
As for the C-examples we collected from~\cite{invgen_dataset}, they originally appear in a substantially different structure and
in a different language than the one we use in our experiments.
Since their task is closer to ours, there is greater concern if
answers for them, i.e., loop invariants, exist online. If so, the LLMs could have leveraged their exposure to these loop invariants by means of transfer-learning. 
Either way, we contend that our reported evaluation results should not be dismissed as memorization.

\subsection{Representativeness of Dataset}

While the models we prompt in the two setups perform very well collectively on our dataset, this dataset may not be representative of real-world industrial examples.
Furthermore, our lemma-mining framework requires the RTL source code in \systemverilog, and is not applicable to netlist-level representations of hardware designs. We are thus unable to assess our current framework on most publicly available circuits.


\subsection{Selection of CoT-Pool}
Our CoT examples were selected manually as a convenience sample, which may introduce selection bias. We chose simple hardware designs from \cite{circuitcove}, a short example from \cite{DBLP:conf/tacas/JainKSC07}, and examples from \cite{invgen_dataset}.
 Most of the CoT examples are taken from~\cite{invgen_dataset}, which is independent of the other benchmark sets, reducing the risk that they are representative.
The example taken from \cite{DBLP:conf/tacas/JainKSC07} is short and structurally simple, and is unlikely to be representative of other designs within this subset. 
For \cite{invgen_dataset}, we explicitly report results on  generalization within this subset and performance on the seen examples, making any potential bias visible. 
\section{Conclusions}

We present a neurosymbolic framework that can be used to automatically generate inductive proofs for hardware verification. While this is only a first step, we believe
such an approach holds promise for industrial settings, where it can assist FV engineers when dealing with complex properties.

In future work, we would like to extend our approach to handle designs with multiple properties simultaneously, allow it to handle larger designs, and evaluate it in an industrial setting.

    

\section*{Acknowledgments}
 The research leading to these results is also partially funded by the Israel Science Foundation (ISF), grant no. 2875/21 and by the European Union (ERC, StrongMC, 101231745). 
Views and opinions expressed are however those of the author(s) only and do not necessarily reflect those of the European Union or the European Research Council Executive Agency. 
Neither the European Union nor the granting authority can be held responsible for them. The authors would also like to thank Amit Eisinger for his help in conducting the rIC3 experiments.




\bibliographystyle{named}
\bibliography{ijcai26}


\end{document}